\documentclass[aps,prl,superscriptaddress,notitlepage,twocolumn]{revtex4-1}
\usepackage{amsmath,amssymb}
\usepackage[utf8x]{inputenc}
\usepackage[english]{babel}
\usepackage{graphicx,epsfig}
\usepackage{txfonts}
\usepackage{xspace}
\usepackage[colorlinks=true,linkcolor=blue,urlcolor=blue,citecolor=blue]{hyperref}
\usepackage{dsfont}
\usepackage{stmaryrd}
\usepackage[usenames,dvipsnames]{color}
\usepackage[normalem]{ulem}

\definecolor{dgreen}{rgb}{0.0, 0.5, 0.0}

\newcommand{\veck}{\mathbf k}

\newcommand{\bed}{\hat b^\dag }
\newcommand{\be}{\hat b }

\definecolor{dgreen}{rgb}{0.0, 0.5, 0.0}

\definecolor{dgreen}{rgb}{0.0, 0.5, 0.0}

\renewcommand{\vec}{\mathbf}

\newcommand{\vecR}{\mathbf R}

\newcommand{\ainv}{1/(n^{1/3}a_\text{IB})}

\global\long\def\ket#1{\left| #1\right\rangle }
 \global\long\def\bra#1{\left\langle #1 \right|}

 \global\long\def\braket#1#2{\left\langle #1\right. \left| #2 \right\rangle }

 \global\long\def\im{\text{Im}}
 \global\long\def\re{\text{Re}}

 \global\long\def\abs#1{\left|#1\right|}

  \global\long\def\v#1{\vec{#1}}

 \global\long\def\arr#1#2#3#4{\left(\begin{array}{cc} #1 & #2 \\ #3 & #4\\ \end{array}\right)}
  
  \global\long\def\col#1#2{\left(\begin{array}{c} #1  \\ #2\\ \end{array}\right)}

\usepackage{array}

\usepackage{bm}

\begin{document}

\title{Quantum dynamics of ultracold Bose polarons}
\author{Yulia E. Shchadilova} 
\affiliation{Department of Physics, Harvard University, Cambridge, Massachusetts 02138, USA}
\author{Richard Schmidt}
\affiliation{Department of Physics, Harvard University, Cambridge, Massachusetts 02138, USA}
\affiliation{ITAMP, Harvard-Smithonian Center for Astrophysics, 60 Garden Street, Cambridge, Massachusetts 02138, USA}
\author{Fabian Grusdt} 
\affiliation{Department of Physics, Harvard University, Cambridge, Massachusetts 02138, USA}
\author{Eugene  Demler}
\affiliation{Department of Physics, Harvard University, Cambridge, Massachusetts 02138, USA}

\date{\today}

\begin{abstract}
We analyze the dynamics of Bose polarons in the vicinity of a Feshbach resonance between the impurity and host atoms. We compute the radio-frequency absorption spectra for the case when the initial state of the impurity is non-interacting and the final state is strongly interacting. We compare results of different theoretical approaches including a single excitation expansion, a self-consistent T-matrix method, and a time-dependent coherent state approach.  Our analysis reveals sharp spectral features arising from metastable states with several Bogoliubov excitations bound to the impurity atom. This surprising result of the interplay of many-body and few-body Efimov type bound state physics can only be obtained by going beyond the commonly used Fr\"ohlich model and including quasiparticle scattering processes. Close to the resonance we find that strong fluctuations lead to a broad, incoherent absorption spectrum where no quasi-particle peak can be assigned.
\end{abstract}

\maketitle

\begin{figure}[t]
\centering
\includegraphics[width=\columnwidth]{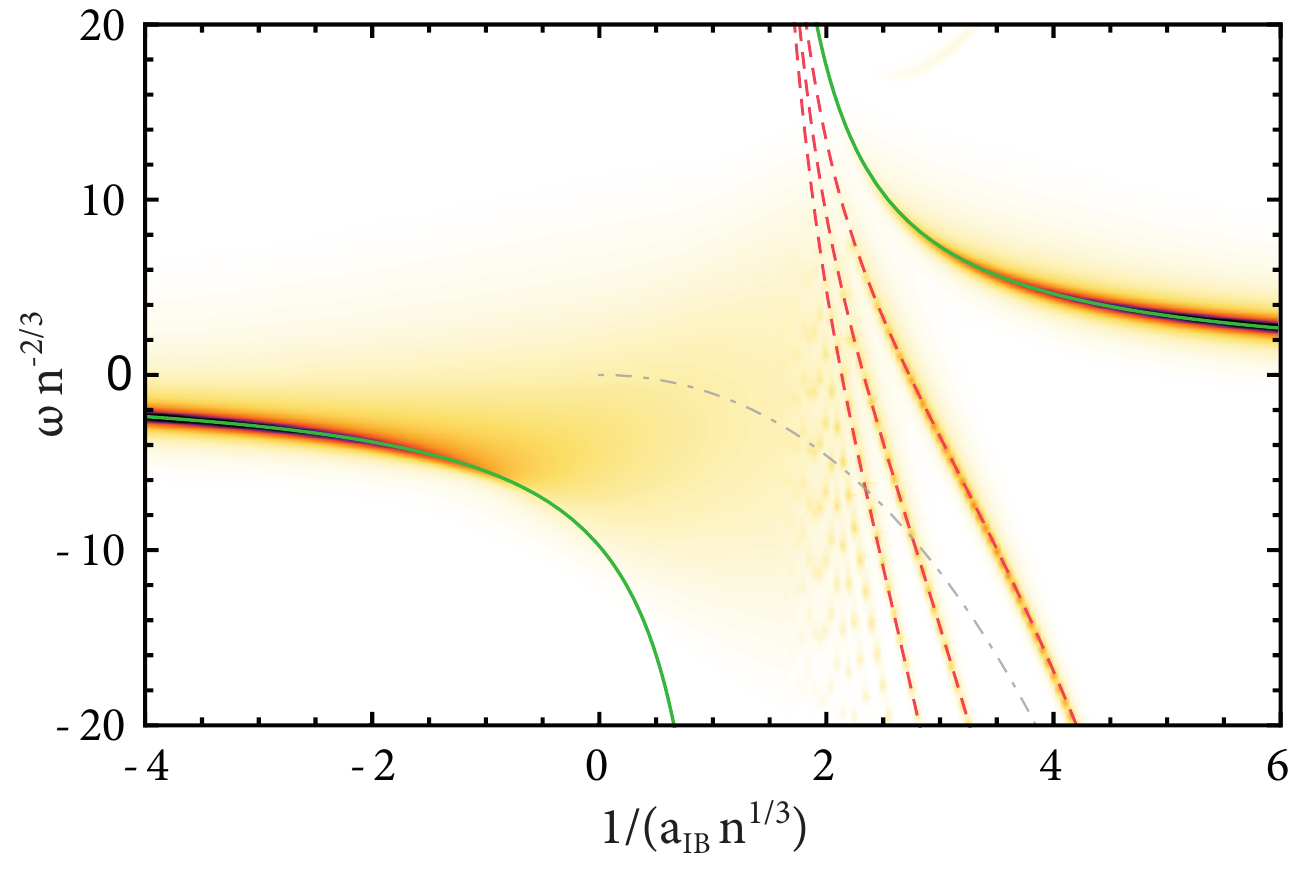}
\vspace{-6mm}
\caption{Absorption spectra $A(\omega)$ of a single impurity immersed in a weakly interacting BEC  as a function of the inverse interaction strength $\ainv$. 
The polaron energy~\eqref{mfenergy} is shown as green line and the energies, cf.~Eq.~\eqref{chevyonMFEnergy}, of  bound states with one, two, and three quasiparticles attached to the polaron are shown as red dashed lines. In the regime of weak repulsion the energy of the first bound state is approaching the dimer binding energy  (gray dash-dotted line). The spectrum is shown for a finite microscopic interaction range corresponding to a momentum cutoff $\Lambda n^{-1/3}=20$ and boson-boson scattering length $a_{BB}n^{-1/3}=0.05$.
}
\label{fig:Spectrum}
\end{figure}

\paragraph{\textbf{Introduction.} --}

Understanding the role of few-body correlations in many-body systems is a challenging problem that arises in many areas of physics. Few particle systems in isolation can be studied using powerful techniques of scattering theory such as Faddeev equations, the hyperspherical formalism, or effective field theory \cite{Pack1987,Fedorov1993,Lin1995,Nielsen2001,glockle2012}. These approaches have been successfully applied to investigate collisions of neutrons and protons \cite{Epelbaum2009,Glockle1996} and Efimov resonances in ultracold atoms \cite{Efimov1970,Braaten2006}. On the other hand the common approach to interacting many-body systems is to use the mean-field approximation, which reduces a many-body problem to  an effective single particle Hamiltonian with self-consistently determined fields. While this approach provides a good description of many fundamental states, including magnetic, superconducting, and superfluid phases \cite{Svistunov}, in many cases it is important to go beyond the mean-field paradigm and include correlations at a few particle level. Recent notable examples include 4e pairing in high Tc superconductors \cite{Berg2009}, spin nematic states \cite{Chubukov1991}, chains and clusters of molecules in ultracold atoms \cite{Wang2006,Dalmonte2011,knap2012cluster,Yao2014}, and the QCD phase diagram  in high-energy physics \cite{Fodor2004,BraunMunzinger2007}. A particularly important class of problems where few-body correlations play a crucial role is the formation of quasiparticles  and polarons. The key feature of both is the dramatic change in the particle dynamics due to the interaction with collective excitations of the many-body system. Famous examples include lattice polarons in semiconductors \cite{Alexandrov1995,Emin2013}, 
 magnetic polarons in strongly correlated electron systems~\cite{Nagaev1975,Trugman1988,Aleksandrov_polaron_book}, and $^3$He atoms in superfluid $^4$He \cite{Bardeen1967}.

Recent experiments with ultracold atoms opened a new chapter in the study of polaronic physics \cite{Chikkatur2000,Palzer2009,Schirotzek2009,Nascimbene2009,Catani2008,kohl,grimm,Zhang2012,Spethmann2012,Balewski2013,Fukuhara2013,Cetina2015,Hohmann2016,Cucchietti2006,Klein2007,Tempere2009,Privitera2010,Novikov2010,Casteels2011pra,Casteels2011lp,Casteels2012pra,Casteels2013,Kain2014,RG,Yin2015,Vlietinck2015,SchmidtLem2015,Christensen2015,Shchadilova2016,SchmidtLem2016}. These systems have tunable interactions between impurity and host atoms \cite{Chin2010} and powerful experimental techniques for characterizing many-body states including spectroscopy \cite{Schirotzek2009,kohl,grimm,Zhang2012}, Ramsey interferometry \cite{Cetina2015}, time of flight experiments \cite{Hohmann2016}, and in-situ measurements with single atom resolution \cite{Bakr2009,Fukuhara2013}. 

In this paper we explore the dynamics of  Bose polarons in the specific setting of radio-frequency (RF) spectroscopy of impurity atoms immersed in a Bose-Einstein condensate (BEC). The most surprising finding of our study is the appearance of sharp spectral lines arising from   states of several Bogoliubov quasiparticles bound to the impurity atom (see Fig.~\ref{fig:Spectrum}). This can be contrasted to earlier theoretical studies that predicted only a single molecular state in addition to attractive and repulsive polarons \cite{Rath2013,Li2014}. Our result demonstrates the importance of the interplay of many-body physics of the BEC of host atoms and few particle Efimov-type physics in the vicinity of a Feshbach resonance  \cite{Song2010,Levinsen2015}. This interplay can not be studied in the commonly used Fr\"ohlich model of BEC polarons because the latter does not include two particle scattering processes that results in the Feshbach resonance. Our work also demonstrates that a broad spectral feature at unitarity can be understood as a superpolaronic state discussed previously in Rydberg systems \cite{SchmidtDem2016,Schlagmueller2016}.

\paragraph{\textbf{Model}. --}
We consider an impurity of mass $m_I$ interacting with a weakly interacting BEC of atoms of mass $m_B$ in the vicinity of an inter-species Feshbach resonance. Within the Bogoliubov approximation the system is described by the Hamiltonian (a derivation is provided in  the Supplementary material (SM))~\cite{Rath2013}, 
\begin{eqnarray}\label{eq:H2}
\hat H&=& g_{\Lambda} n + \frac{\hat{\v P}^2 }{2m_I} 
+\sum_{\v{k}} \omega_{\v{ k}}\hat{b}_{\v k}^{\dagger}\hat{b}_{\v k} + 
\nonumber\\ &&
 \frac{g_{\Lambda}\sqrt{n}}{L^{d/2}} \sum_{\v k} W_{\v k}e^{i\v k \hat{\v R}}\left(\hat{b}_{\v k}^{\dagger} + \hat{b}_{-\v k}\right)
+\frac{ g_{\Lambda}}{L^d} \sum_{\v k\v k'} V^{(1)}_{\v k\v k'}  e^{i\v (k-k') \hat{\v R}} \hat b_{\v k}^\dag \hat b_{\v k'} 
\nonumber\\ &&
+  \frac{g_{\Lambda}}{L^d} \sum_{\v k\v k'} V^{(2)}_{\v k\v k'} e^{i\v (k+k') \hat{\v R}} \left(  \hat b_{\v k}^\dag \hat b_{\v k'}^\dag + \hat b_{-\v k} \hat b_{-\v k'}\right).
\end{eqnarray}
Here the operators  $b^\dag_{\v k}$ ($b_{\v k}$) create (annihilate) Bogoliubov quasiparticles (`phonons') with momentum $\v k$ and dispersion $ \omega_{\v k} $. The bare inter-species interaction is given by $g_\Lambda$. Furthermore $W_{\v k}= \sqrt{\varepsilon_{\v k}/\omega_{\v k}}$, and  $V_{\v k\v k'}^{(1)} \pm V^{(2)}_{\v k \v k'}= \left( W_{\v k} W_{\v k'}\right)^{\pm 1}$ define the interaction vertices where $\varepsilon_{\v k}= k^2/2 m_B$ is the dispersion relation of bare host atoms; $n$ is the condensate density, and $L^d$ the system's volume. 

The last two lines in Eq.~\eqref{eq:H2} describe the interaction of the impurity at position $\hat \vecR$ and momentum $\hat{\v P}$ with the host bosons. In the  Fr\"ohlich model only the interaction term linear in the bosonic operators is present and it describes the creation of excitations directly from the BEC. However,  a microscopic derivation  reveals that in cold atomic systems also the additional quadratic terms, included in Eq.~\eqref{eq:H2}, are present which  lead to rich physics beyond the Fr\"ohlich paradigm \cite{Rath2013}.

The extended Hamiltonian  \eqref{eq:H2} allows for a proper regularization of the contact interaction between the impurity and bosons which is beyond the Fr\"ohlich model.
From the solution of the two-body scattering problem of Eq.~\eqref{eq:H2} follows the relation of the microscopic interaction strength $g_\Lambda$ to the impurity-boson scattering length $a_{IB}$ by the Lippmann-Schwinger equation 
\begin{equation}\label{eq:LippSchwing}
g_\Lambda^{-1} =  \frac{\mu_{\rm red}}{2\pi} a_{IB}^{-1} -\frac{1}{L^d} \sum_{\v k}^\Lambda \frac{2\mu_\text{red}}{\veck^2}.
\end{equation}
Here $\mu_{\rm red}=m_I m_B/(m_I+m_B)$ is the reduced mass and $\Lambda\sim1/ r_0$ denotes an ultraviolet (UV) cutoff scale  related to a finite range $r_0$ of the interaction potential. In the limit $\Lambda\to\infty$  contact interactions are recovered. We note that all our numerical results are obtained for the mass-balanced case $m_I=m_B$.

We describe the impurity-bath system in the frame co-moving with the polaronic quasiparticle \cite{Lee1953}. This is achieved using a canonical transformation $\hat{\mathcal H}=\hat S^{-1}\hat H \hat S$ with $\hat S=e^{i \hat{\vec R} \hat{ \vec P}_B}$ where $\hat{\vec P}_B=\sum_\veck \veck \bed_\veck\be_\veck$ is the total momentum operator of the bosons. After the transformation sectors with different total system momentum $\vec{P}$ are decoupled in the Hamiltonian  $\hat {\mathcal H}$. 
The bosons now interact with each other since the impurity kinetic energy transforms according to $\hat {\v P}^2/2M \to (\hat {\v P} - \hat{\vec P}_B)^2/2M$ \cite{Shashi2014,SchmidtLem2016,sup}, for details see SM.

\paragraph{\textbf{Quantum quench dynamics}. --} In this work we predict the excitation spectrum of Eq.~\eqref{eq:H2}. In experiments the spectrum can be explored using RF spectroscopy where the impurity is initially in a spin-state $\ket{\downarrow}$ in which it does not interact with the bath. Then the response is measured with respect to a weak monochromatic perturbation $e^{i\omega t}\ket{\uparrow}\bra{\downarrow}+h.c.$  flipping the spin to a state $\ket{\uparrow}$ which does interact with the bosons with a strength determined by the dimensionless interacting parameter $1/(n^{-1/3} a_{IB})$. Using Fermi's golden rule we calculate the absorption spectrum -- i.e.~the impurity spectral function --  in linear response $A(\omega)=2\,\text{Re}\int_0^\infty dt e^{i\omega t}S(t)$, where $S(t)=\bra{\Psi(0)}e^{-i \hat H t}\ket{\Psi(0)}$ is  a time-dependent overlap.
Here $\ket{\Psi(0)}$ denotes the initial state of the system and the overlap $S(t)$ describes the dynamics of the system after a quench of the interactions between impurity and the bath. In real-time the overlap $S(t)$ can be measured using Ramsey interferometry \cite{Knap2012,Cetinainprep}.

\begin{figure}[t]
\centering
\includegraphics[width=\columnwidth]{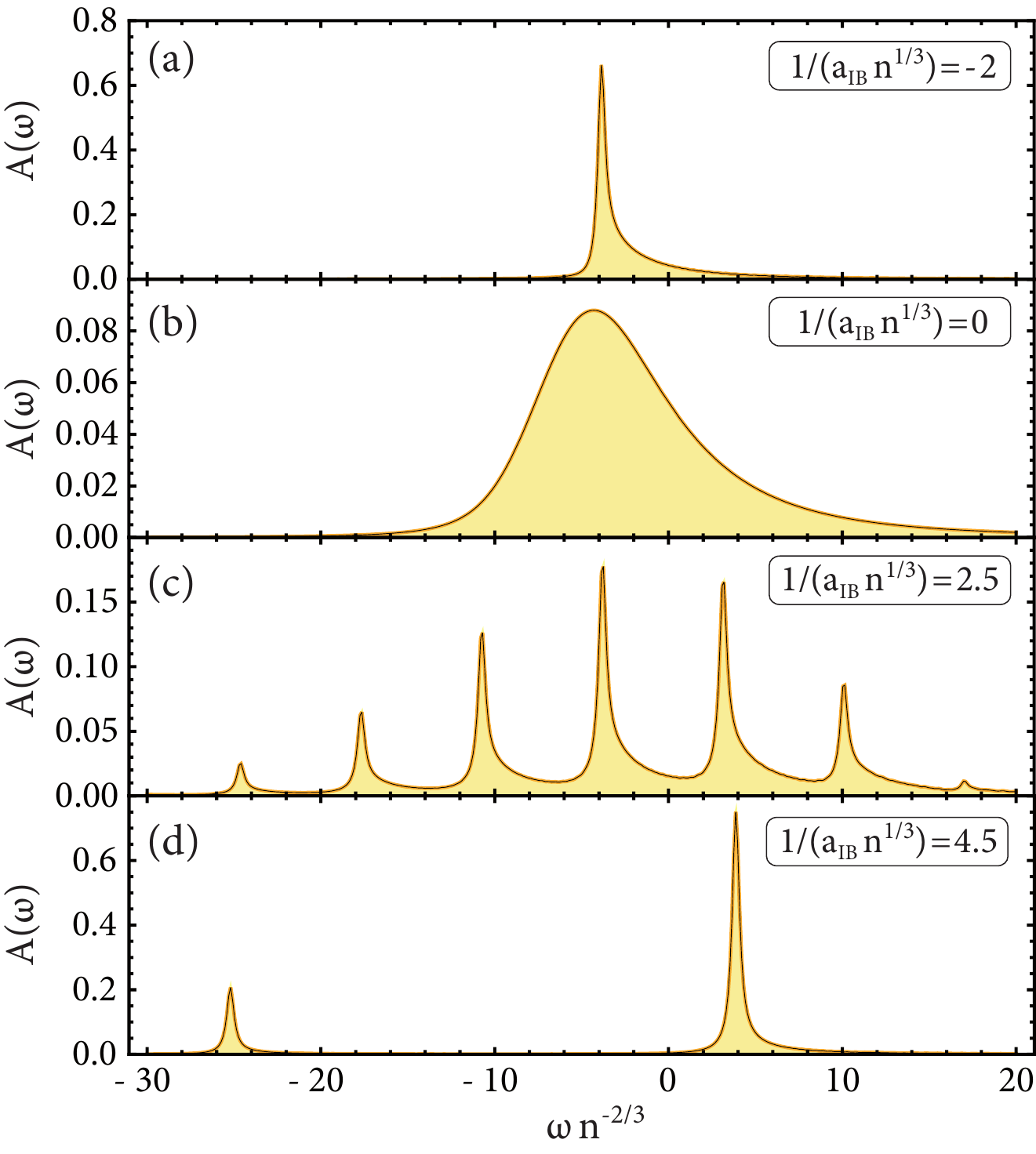}
\vspace{-6mm}
\caption{Absorption spectra $A(\omega)$ of a single impurity immersed in a weakly interacting Bose gas for the fixed scattering length $\ainv$: (a) $\ainv=-2$ , (b) $\ainv=0$, (c) $a_{IB}^{-1}=2.5$, (d) $\ainv=4.5$.
For the negative scattering length, panel (a), the absorption spectrum reveals the attractive polaron which follows closely the  polaron energy~\eqref{mfenergy}.  
At unitarity, panel (b), the spectrum shows a broad spectral feature and no quasi particle peak can be assigned. 
As the Feshbach resonance is crossed towards positive scattering lengths, panel (c), a series of bound states emerges.  
For small positive scattering length, panel (d), the binding energy of the negative frequency peak increases while at positive frequencies a long-lived repulsive polaron becomes the dominant excitation. As for Fig.~\ref{fig:Spectrum} we artificially broadened the spectrum to make sharp features visible.}
\label{fig:SpectrumCuts}
\end{figure}

In order to predict the real-time evolution as well as the excitation spectrum of the system, we invoke the time-dependent variational principle \cite{Jackiw1979}. The approach relies on a projection of the many-body wave function onto a submanifold of the full Hilbert space spanned by a set of trial wave functions. 

Specifically, we employ a variational state in the form of a product of coherent states~\cite{Shashi2014}
\begin{equation} \label{eq:WF}
\ket{\Psi_\textrm{coh}(t)} = e^{-i \phi(t)} e^{\sum_{\v k} \beta_{\v k}(t) b_{\v k}^\dag - h.c. } \ket{0}
\end{equation}
where $\beta_{\v k }(t)$ are the coherent amplitudes, $\phi(t)$ is a global phase which ensures energy conservation, and $\ket{0}$ denotes the  vacuum of Bogoliubov quasiparticles.
The ansatz \eqref{eq:WF} provides an exact solution when describing the sudden immersion of an impurity of infinite mass into a gas of non-interacting bosons.
As such the choice of the wave function is based  on an exact limit of the model which is valid for arbitrary interaction strength between the impurity and Bose gas.

Before turning to the dynamics, we consider the parameters in Eq.~\eqref{eq:WF} as time-independent variables and study  the variational energy $\bra{\Psi_\textrm{coh}}\hat{\mathcal H}\ket{\Psi_\textrm{coh}}$ of the system.
Its variation yields the polaron energy which for  zero total momentum, $\vec{P}=0$, reads \cite{sup} 
\begin{eqnarray} \label{mfenergy}
E_{\rm pol} &=& \frac{2\pi}{\mu_{\rm red}}\frac{n}{a_{IB}^{-1} - a_0^{-1}}
\end{eqnarray}
where $a_0^{-1} = \frac{2\pi}{\mu_{\rm red}} \sum_{\v k} \left( 2\mu_{\rm red}/ \veck^2 - W_{\v k}^2/(\omega_{\v k} + \v k^2/2m_I) \right)$ defines the shift of the scattering resonance due to the many-body environment. We note that the expression~\eqref{mfenergy} correctly accounts for the regularization of the short-range interaction as given by Eq.~\eqref{eq:LippSchwing} and, accordingly,  contains only experimentally accessible parameters.
The attractive and repulsive polaron branches described by Eq.~\eqref{mfenergy} are shown in Fig.~\ref{fig:Spectrum} and play an important role in the dynamics as we describe below.

For  the time-dependent problem we promote the variational parameters $\beta_\veck(t)$ and $\phi(t)$ to time dependent quantities. The equation of motions $\frac{d}{dt}\frac{\partial \mathcal L}{\partial \dot \beta}-\frac{\partial \mathcal L}{\partial \beta}=0$ for the  parameters $\beta_{\veck}$ are obtained from the Lagrangian $\mathcal L=\bra{\Psi_\textrm{coh}}i\partial_t-\hat{\mathcal H}\ket{\Psi_\textrm{coh}}$ of the system. They can be cast in the form 
\begin{eqnarray}\label{Dyn}
i \dot \beta_{k} &=& g_{\Lambda} \sqrt{n} W_{\v k} +\left( \omega_{\v k} + \frac{\v k^2}{2m_I} - \frac{\v k \left( {\v P} - \v P_B \left[ \beta_{\v k}\right]\right)}{m_I}\right) \beta_{\v k} \nonumber\\    
&+& \frac{g_{\Lambda}}{2} \left[W_{\v k}\sum_{\v k'} W_{\v k'}\left(  \beta_{\v k'}+  \beta_{\v k'}^*\right)  
+ W_{\v k}^{-1}\sum_{\v k'} W_{\v k'}^{-1}\left(  \beta_{\v k'}-  \beta_{\v k'}^*\right) \right]  \nonumber\\ 
\dot \phi(t) &=& g_{\Lambda} n + \frac{1}{2} g_{\Lambda}\sqrt{n} \sum_{\v k} W_{\v k} \left( \beta_{\v k} + \beta_{\v k}^*\right) +\frac{{\v P}^2 -\v P_B^2 \left[ \beta_{\v k}\right]}{2m_I}
\end{eqnarray}
where $\v P_B \left[ \beta_{\v k}\right]=\sum_{\v k} \v k \abs{\beta_{\v k}}^2$ is the total momentum of the phonons. In Eq.~\eqref{Dyn} $g_\Lambda$ is defined by the Lippman-Schwinger equation~\eqref{eq:LippSchwing} with given scattering length $a_{IB}$ and a finite UV cutoff $\Lambda$. 
This ensures a time-evolution which is fully regularized and free of any divergencies. 

\paragraph{\textbf{Absorption spectra}.--} In the class of  coherent states, Eq.~\eqref{eq:WF}, the dynamical overlap becomes
$S(t) =\braket{0}{\Psi_\textrm{coh}(t)}= \exp\left[{-i\phi(t)-\frac{1}{2} \sum_{\v k } \abs{\beta_{\v k}(t)}^2}\right]$.
From its Fourier transform we obtain the rf absorption spectrum, shown in
Fig.~\ref{fig:Spectrum}, as function of the inverse interaction strength $1/(n^{1/3}a_\text{IB})$.

The spectrum exhibits two main excitation branches: the attractive polaron for $a_{IB}<0$ and the repulsive polaron for $a_{IB}>0$. Both follow the energy Eq.~\eqref{mfenergy} (indicated as green lines in Fig.~\ref{fig:Spectrum}), and in the weak coupling regime both are well described by a Fr\"ohlich polaron model with renormalized interaction parameters \cite{Rath2013}. 

The attractive polaron is formed when the impurity is dressed by bosonic excitations due to the weak attractive interactions with the bath for $a_{IB}<0$. The corresponding spectral signature, shown in Fig.~\ref{fig:SpectrumCuts}(a) for $1/(n^{1/3}a_\text{IB})=-2$, is a sharp quasiparticle peak at negative frequencies given by  Eq.~\eqref{mfenergy}. 

As the Feshbach resonance at $\ainv=0$ is approached the attractive polaron peak looses spectral weight to the scattering continuum at higher frequencies. 
Close to unitarity, no particular eigenstate of $\hat{\mathcal H}$  yields a distinct contribution to the dynamical overlap $S(t)$, and many overlaps between the eigenstates of the many-body Hamiltonian and the non-interacting state of the system are of the same order. Hence the spectrum becomes broad and no coherent quasiparticle excitation is possible any longer. In consequence, perturbative approaches based on expansions around the non-interacting state become particularly unreliable in this strong coupling regime.

For positive scattering length, i.e.~$\ainv>0$, the effective interaction between the mobile impurity and the bosons is repulsive and leads to the formation of a repulsive polaron. This  state manifests itself as a quasiparticle peak at positive frequency.  As can be seen in Fig.~\ref{fig:Spectrum}, the energy of the repulsive polaron increases as the shifted resonance at $a_{IB}=a_0$ is approached, and  it follows the saddle point prediction Eq.~\eqref{mfenergy}.
Similar to the attractive polaron, close to resonance, the repulsive polaron quickly looses quasiparticle weight.

However, the spectral weight is transferred not only to incoherent excitations but also to coherent spectral features which appear below the repulsive polaron branch. In Fig.~\ref{fig:Spectrum} and Fig.~\ref{fig:SpectrumCuts}(c) those features are visible as a series of equidistant peaks.
Such excitations -- corresponding to multiple bosons bound to the impurity -- are absent in the  Fr\"ohlich model since they are a consequence of strong pairing correlations which originate from the quadratic interaction terms in Eq.~\eqref{eq:H2}. As unitarity is approached, the spacing between these bound state peaks decreases until they eventually cannot be resolved. We note that such a crossover in the spectral profile from discrete bound state levels to a broad distribution is reminiscent of  the formation of superpolarons which has been studied in Rydberg molecular systems  \cite{SchmidtDem2016,Schlagmueller2016}. 

\paragraph{\textbf{Many-body bound states.} --}

The emergence of the series of bound states on the repulsive side  of the resonance ($1/a_\text{IB}>0$) is a novel feature of impurities immersed in atomic BECs. Each peak corresponds to a single, two, and more Bogoliubov quasiparticles bound to the repulsive polaron.

The structure of the bound state spectrum can be understood analytically. We consider a wave-function which accounts for a single Bogoliubov excitation above the polaron state $\ket{\Psi_\text{pol}}$ \cite{SchmidtLem2016}
\begin{equation}\label{chevyonMF}
\ket{\Psi'(t)}=\sum_\veck \gamma_\veck(t) \bed_\veck \ket{\Psi_\text{pol}}
\end{equation} 
The ansatz Eq.~\eqref{chevyonMF} can be regarded as a molecular wave function which fully accounts for two-body bound state physics on top of the repulsive polaron state $\ket{\Psi_\text{pol}}$. A  calculation shows that the equations of motion of this state have  an eigenmode $\gamma_\veck(t)\sim e^{-i\omega t}$ with eigenfrequency $\omega$ determined by the solution of the equation (a derivation is given in the SM \cite{sup})
\begin{equation}\label{chevyonMFEnergy}
\frac{\mu_\text{red}}{2\pi a_{IB}}-\sum_\veck \left[\frac{\left(W_\veck^2+W_\veck^{-2}\right)/2}{\omega-E_\text{pol}-\Omega_\veck}+\frac{2\mu_\text{red}}{\veck^2}\right]=0
\end{equation}
where $\Omega_\veck= \omega_\veck + \v k^2/2m_I $. In Fig.~\ref{fig:Spectrum} we show the energy $\omega$ as a dashed red lines. The lines occur in integer multiples of $\omega$ as the bound state can be occupied by several phonons at the same times, an effect  taken into account by the exponentiated creation operators in Eq.~\eqref{eq:WF}.

For a further understanding of these states we first consider the limit of the BEC density going to zero, ie.~$n\to 0$, $\ket{\Psi_\text{pol}}\to \ket{0}$, $W_\veck\to 1$ and $\Omega_\veck\to\veck^2/2\mu_\text{red}$. 
This limit defines the two-body problem where, for zero-range interactions, the bound state energy becomes $\epsilon_B=-\hbar^2/\mu_\text{red}a_{IB}^2$ \cite{Rath2013}, which is fully recovered by Eq.~\eqref{chevyonMF}.
For any finite density of non-interacting bosons, and assuming an infinitely heavy impurity, this bound state can be occupied by arbitrarily many bosons, and each bound atom contributes an energy $\epsilon_B$. In an interacting Bose gas, the {infinitely massive} impurity scatters with Bogoliubov quasiparticles instead of bare bosons. As a consequence the binding energy $\epsilon_B$ is modified. Since in the Bogoliubov model quasiparticles do not interact with each other they can still occupy the bound state multiple times which leads to the series of spectral lines visible in Fig.~\ref{fig:SpectrumCuts}(c).

The emergence of the sequence of many-body bound states revealed in our approach is related to the Efimov effect \cite{Efimov1970,Braaten2006,Schmidt2012}. Indeed, an exact solution of the corresponding three-body problem reveals that, as the infinite mass condition is relaxed, recoil leads to the splitting of three-, four-, etc. body bound states into an infinite series of bound states situated in a regime exponentially close to the Feshbach resonance \cite{Efimov1972,Stecher2009}. Within our approach this splitting is absent since the effective interactions between phonons vanishes for the class of coherent states Eq.~\eqref{eq:WF}. In consequence, in the limit of vanishing density, we effectively recover the Efimov physics  of an infinitely mass-imbalanced system, as discussed by Efimov in his seminal work \cite{Efimov1972,Efimov1973}.

\paragraph{\textbf{Comparison to other approaches}.--}

As can be seen from the previous analysis, a simple ansatz such as Eq.~\eqref{chevyonMF} can already account for aspects of  complex many-body bound state physics. Indeed, the ansatz~\eqref{chevyonMF}  is related to a variational wave function which is based to an expansion in terms of single particle excitations. In the context of cold atoms, it was first introduced in the work by Chevy for fermionic systems \cite{chevy_universal_2006}, and later generalized to bosonic systems \cite{Rath2013,Li2014,SchmidtLem2015,SchmidtLem2016}. 

To connect our results to previous work we present here an extension of the equilibrium approach to real-time dynamics by studying the time-dependent variational wave function
\begin{equation} \label{ChevyFull}
\ket{\Psi_\text{1ex}(t)}=\alpha_0(t)\ket{0}+\sum_\veck \alpha_\veck(t) \bed_\veck \ket{0}
\end{equation}
This wave-function accounts for a single phonon excitation on top of the unperturbed BEC state.
We calculate the excitation spectrum of the system from the dynamical overlap $S(t) = \alpha_0(t)$ as obtained from the equation of motions of the variational parameters. This ansatz is naturally connected to the coherent state approach as an expansion in low occupation numbers, which justifies the validity of Eq.~\eqref{ChevyFull} in the limit of low densities.  While in the weak coupling regime the simple ansatz \eqref{ChevyFull} reproduces the predictions using the coherent state approach,  it fails to describe the intricate many-body physics in regimes where multiple boson excitations become relevant (for a comparison see the SM \cite{sup}).

\begin{figure}[t]
\centering
\includegraphics[width=\columnwidth]{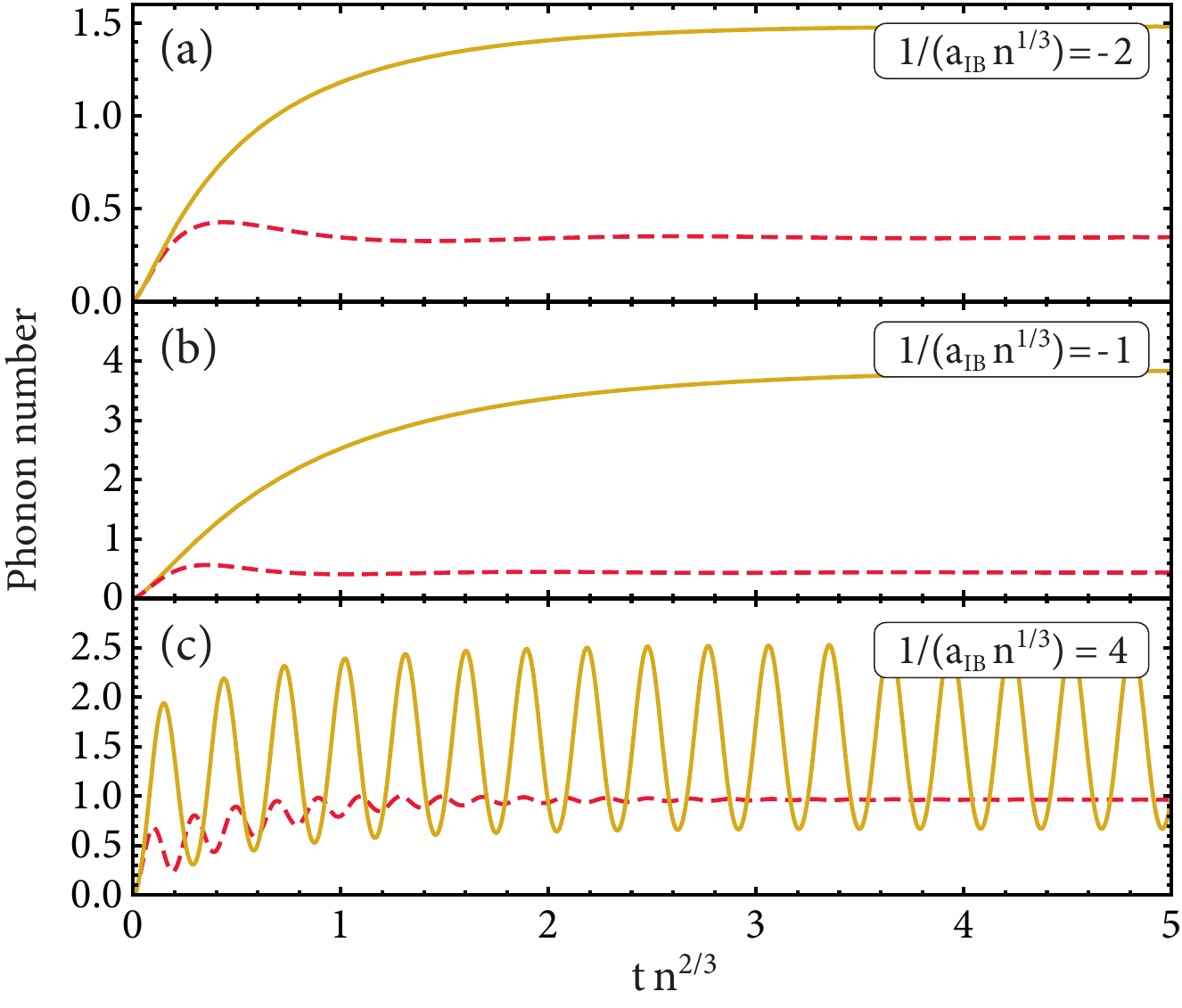}
\caption{Time evolution of the number of phonons, $N_\textrm{ph}$, for attractive and repulsive interactions, (a) $\ainv = -2 $, (b) $\ainv = -1$, (c) $\ainv= 4$, obtained by the coherent state approach~\eqref{eq:WF} and the single-excitation expansion~\eqref{ChevyFull} (yellow solid and red dashed lines, respectively).}
\label{fig:PhNum}
\end{figure}

The difference between the coherent states approach~\eqref{eq:WF} and the single-excitation expansion~\eqref{ChevyFull}
can be highlighted when the time evolution of the number of phonon excitations $N_\textrm{ph}=\langle\sum_{\v k} \hat b_{\v k}^\dagger \hat b_{\v k}\rangle$ is compared, where $\langle \ldots\rangle$ denotes the expectation value with respect to the chosen variational state. In Fig.~\ref{fig:PhNum} we show $N_\textrm{ph}(t)$  for attractive and repulsive interactions. 
On the attractive side both approaches predict a saturation of the phonon number in the long time limit, see panels (a) and (b) in Fig.~\ref{fig:PhNum}. In the single-excitation expansion (dashed lines), the number of excitations is restricted to one. This limitation of Eq.~\eqref{ChevyFull} becomes apparent when comparing it to the coherent state approach (solid lines).  We find that already for moderate attractive coupling strength the number of phonons exceeds one. Hence, due to the restriction in the number of  excitations for the ansatz~\eqref{ChevyFull}, both approaches agree only in the initial short time evolution. 

On the repulsive side the coherent state approach predicts oscillations of the number of phonons in the long time limit, see panel (c) in Fig.~\ref{fig:PhNum}. These oscillations appear due to the competition between the many-body polaron branch and the few-body bound states in real time. 
In contrast, in the single-excitation expansion these oscillations decay gradually and in the long time limit, the number of excitations saturates at one, reflecting the formation of a single molecular state. 

We note that similar to the time-dependent coherent state approach presented in this letter, a self-consistent T-matrix approach \cite{Rath2013} also accounts for an infinite number of bosonic excitations. Using this  diagrammatic approach  it had been found  that the inclusion of multiple boson excitations has a profound influence on the spectrum. 
However multiphonon bound states have not been observed in this earlier work.
Yet we emphasize that when considering only moderate interaction strengths or short-time evolution, expansions in terms of a few particle excitations such as Eq.~\eqref{ChevyFull} remain a viable approach. They correctly describe the weakly attractive and repulsive polaron branches as well as the one-boson bound state present in the spectrum sufficiently far away from the Feshbach resonance.

\paragraph{\textbf{Summary and Outlook}.--}

In summary, we analyzed the dynamics and absorption spectra of an impurity immersed in a BEC. We demonstrated both the disappearance of the sharp quasiparticle spectral feature at strong coupling and the presence of a novel type of excitations in which several Bogoliubov quasiparticles are bound to the impurity particle.  Our analysis highlights the importance of quasiparticle scattering processes that are not present in the commonly used Fr\"ohlich model. They result in strong short distance correlations that give rise to Efimov type physics of multi-particle bound states and play a crucial role in suppressing the quasiparticle spectral weight close to the Feshbach resonance. Our work opens new directions for studying non-perturbative phenomena in Bose polarons at strong coupling. We expect that new insight into such systems can be gained by extending our analysis to more sophisticated approaches such as Gaussian variational wave functions~\cite{Shchadilova2016} and Renormalization Group analysis~\cite{RG}. Our method can also be extended to problems beyond linear response such as Rabi oscillations of strongly driven Bose impurities~\cite{grimm,Knap2013}.

\paragraph{\textbf{Acknowledgment.} -- } 
We thank Dima Abanin,  Gregory Astrakharchik,  Immanuel Bloch, Ignacio Cirac, Thierry Giamarchi,  Christoph Gohle, Deborah Jin, Marton Kanasz-Nagy, Dries Sels, Tao Shi, Lars Wacker, and Martin Zwierlein for inspiring discussions.
The authors acknowledge support from the NSF Grant No.~DMR-1308435, Harvard-MIT CUA, AFOSR New Quantum Phases of Matter MURI, the ARO-MURI on Atomtronics, ARO MURI Quism program. E.~D. acknowledges support from the Simons foundation, the Humboldt Foundation, Dr.~Max~R\"ossler, the Walter Haefner Foundation, and the ETH Foundation. R.~S. is supported by the NSF through a grant for the Institute for Theoretical Atomic, Molecular, and Optical Physics at Harvard University and the Smithsonian Astrophysical Observatory. F.~G. is grateful for financial support from the Gordon and Betty Moore foundation.

\vspace{1cm}
\begin{center}
\Large{Supplementary Material}
\end{center}

\section{I. Derivation of the effective model}
We consider the microscopic model which describes the interaction of a single particle of mass $m_I$ with a weakly interacting  gas of bosons of  mass $m_B$. We assume that the interactions between the particles are described by contact interactions.
This approximation is  valid  for the description of ultracold gases close to a broad inter-species Feshbach resonance \cite{Rath2013}. The Hamiltonian reads
\begin{multline}\label{eq:H1}
\hat H = \sum_{\v k} \epsilon_{\v k}^B \hat{a}_{\v k}^\dag \hat{a}_{\v k} + \frac{g_{BB}}{2} \sum_{\v k \v k' \v q} \hat{a}_{\v k+\v q}^\dag \hat{a}_{\v k'-\v q}^\dag \hat{a}_{\v k}  \hat{a}_{\v k'}\\ 
+\sum_{\v k} \epsilon_{\v k}^I \hat{d}_{\v k}^\dag\hat{d}_{\v k} +
g_{\Lambda}\sum_{\v k \v k' \v q} \hat{a}_{\v k+\v q}^\dag  \hat{a}_{\v k}  \hat{d}_{\v k'-\v q}^\dag \hat{d}_{\v k'}
\end{multline}
where $\hat{a}_{\v k}^\dag$ and $\hat{d}_{\v k}^\dag$  are the creation operators that describe the bosonic host particles and an impurity with the corresponding  dispersion relations $\epsilon^B_{\v k} =\frac{k^2}{2 m_B}$ and $\epsilon^I_{\v k} =\frac{k^2}{2 m_I}$. The parameters of the contact interaction (bare interaction) are given by the coefficient  $g_{\Lambda}$  for the interaction between the host particles and the impurity and $g_{BB}$ for the interaction among the host particles. We allow for the general case of a mass imbalanced system where the mass of the bosonic field $m_B$ and the mass of the impurity $m_I$ are not fixed to be equal. 

We account for the macroscopic occupation of the $\v k =0$ mode within the Bogoliubov approximation \cite{Pitaevskii2003}. Following the corresponding truncation of the Hamiltonian on the quadratic level in the bosonic creation and annihilation operators we diagonalize the purely bosonic part of the Hamiltonian by a unitary Bogoliubov rotation\begin{equation}
a_{\v k} = u_{\v k} b_{\v k} - v_{\v k} b_{-\v k}^\dag.
\end{equation}
Here the coefficients are given by $u_\mathbf{k}^2 = \frac{1}{2} \left( \frac{\varepsilon_{\v k}+g_{\rm BB} n}{\omega_{\v k}}+1\right)$, $v_\mathbf{k}^2=\frac{1}{2} \left( \frac{\varepsilon_{\v k}+g_{\rm BB} n}{\omega_{\v k}}-1\right)$. This transformation  introduces  the Bogoliubov quasiparticles with dispersion relation $\omega_{\v k}$ which are created by the operators $b_{\v k}^\dag$. Reexpressing the resulting Hamiltonian in explicit impurity coordinates yields  the Hamiltonian~\eqref{eq:H2} in the main text.

In order to take advantage of  momentum conservation we perform a canonical transformation to the so-called polaron frame using $\hat S=e^{i \hat{\vec R} \hat{ \vec P}_B}$ where $\hat{\vec P}_B=\sum_\veck \veck \bed_\veck\be_\veck$ is the total momentum operator of the bosons. The operator $\hat S$ transforms the momentum of the impurity and bosonic operators in the following way
\begin{eqnarray}
\hat S^{-1} \hat{\v P} \hat S &=& \hat{\v P} - \sum_\veck \veck \bed_\veck\be_\veck \\ \nonumber
\hat S^{-1} \hat b_{\v k} \hat S &=& \hat b_{\v k} e^{-i \hat{\vec R} \v k }.
\end{eqnarray}
Applying this transformation to the Hamiltonian~\eqref{eq:H2}, $\hat{\mathcal H} =\hat S^{-1}\hat H \hat S$,  we obtain
\begin{eqnarray}\label{eq:H3}
\hat{\mathcal H}&=&=  g_{\Lambda} n + \frac{1 }{2m_I} \left( \hat{\v P} - \sum_\veck \veck \bed_\veck\be_\veck \right)^2
+\sum_{\v{k}} \omega_{\v{ k}}\hat{b}_{\v k}^{\dagger}\hat{b}_{\v k} + 
\nonumber\\ &&
 \frac{g_{\Lambda}\sqrt{n}}{L^{d/2}} \sum_{\v k} W_{\v k} \left(\hat{b}_{\v k}^{\dagger} + \hat{b}_{-\v k}\right)
+\frac{ g_{\Lambda}}{L^d} \sum_{\v k\v k'} V^{(1)}_{\v k\v k'} b_{\v k}^\dag b_{\v k'} 
\nonumber\\ &&
+  \frac{g_{\Lambda}}{L^d} \sum_{\v k\v k'} V^{(2)}_{\v k\v k'} \left(  b_{\v k}^\dag b_{\v k'}^\dag + b_{-\v k} b_{-\v k'}\right).
\end{eqnarray}
Since $\hat{\v P}$ now commutes with the Hamiltonian,  the Hilbert space of the Hamiltonian $\hat{\mathcal H}$~\eqref{eq:H3} becomes a direct product of the sub-spaces characterized by the total momentum of the system.

\section{II. Derivation of the saddle point solution}
In this section we derive the saddle point solution of the equations of motions of the parameters in Eq.~\eqref{Dyn}. 
We set the left hand side of the first equation to zero and obtain the following equation which defines the saddle point
\begin{eqnarray}
0&=&g_{\Lambda} \sqrt{n} W_{\v k} + \Omega_{\v k} \beta_{\v k}  \\ \nonumber
&+& \frac{1}{2} g_{\Lambda} W_{\v k}\sum_{\v k'} W_{\v k'}\left(  \beta_{\v k'}+  \beta_{\v k'}^*\right) 
+\frac{1}{2} g_{\Lambda} W_{\v k}^{-1}\sum_{\v k'} W_{\v k'}^{-1}\left(  \beta_{\v k'}-  \beta_{\v k'}^*\right).  
\end{eqnarray}
Here  $\Omega_{\v k}=\omega_{\v k} +\frac{{\v k}^2}{2M} - \frac{\v k}{M} (\v P- \v P_{B})$, $\v P_{B} = \sum_{\v k} \v k \abs{\beta_{\v k}}^2$, and $\vec P$ denotes the total system momentum.
The real and imaginary parts of the coherent amplitudes $\beta_{\v k}$ can be expressed as
\begin{eqnarray}\label{MF_ampl}
\re\beta_{\v k} &=& -g_{\Lambda} W_{\v k} \frac{  \sqrt{n} +\sum_{\v k'} W_{\v k'} \re\beta_{\v k'} }{\Omega_{\v k}}, \\ \nonumber
\im\beta_{\v k} &=& -g_{\Lambda} W_{\v k}^{-1} \frac{  \sum_{\v k'} W_{\v k'}^{-1} \im\beta_{\v k'} }{\Omega_{\v k}}.
\end{eqnarray}

The first equation allows for an analytical solution which is found by a multiplication with $W_{\v k}$, a summation over all modes and expressing the problem in terms of the averaged quantity $\chi=\sum_{\v k} W_{\v k} \re\beta_{\v k}$.
After some algebra one obtains
\begin{eqnarray} \label{MF_eq_2}
&&  \re \beta_{\v k}=- \frac{\sqrt{n} W_{\v k}}{\Omega_{\v k}} \frac{1}{ g_{\Lambda}^{-1}+\sum_{\v k} \frac{W_{\v k}^2}{\Omega_{\v k}}}.
\end{eqnarray}
The microscopic interaction strength $g_\Lambda$ is related to the impurity-boson scattering length $a_{IB}$ by the zero-momentum limit of the Lippmann-Schwinger equation which yields
\begin{equation}
a_{IB}^{-1} =  \frac{2\pi}{\mu_{\rm red}} g_\Lambda^{-1} + 4\pi \sum_{\v k}^\Lambda \frac{1}{\veck^2}
\end{equation}
where $\Lambda$ denotes a high-momentum cutoff which can be taken to infinity.
Upon insertion into Eq.~\eqref{MF_eq_2} one obtains 
\begin{eqnarray}\label{MF_final_app}
&&  \re \beta_{\v k}=- \frac{2\pi}{\mu_{\rm red}} \frac{1}{ a_{IB}^{-1}-a_+^{-1}}\frac{\sqrt{n} W_{\v k}}{ \omega_{\v k} +\frac{{\v k}^2}{2M} - \frac{\v k}{M} (\v P- \v P_{B})}
\end{eqnarray}
where we defined $a_+^{-1}$ as
\begin{eqnarray}\label{a0_app}
a_+^{-1} &=& \frac{2\pi}{\mu_{\rm red}} \sum_{\v k} \left( \frac{2\mu_{\rm red}}{ \v k^2} - \frac{W_{\v k}^2}{ \omega_{\v k} +\frac{{\v k}^2}{2M} - \frac{\v k}{M} (\v P- \v P_{B})} \right).
\end{eqnarray}
The second equation has the trivial solution $\im\beta_{\v k} =0$.

After the coherent amplitudes are expressed in terms of the inverse scattering length, a self-consistent procedure for the determination of  $a_+^{-1}$ and the total momentum of the phonons can be set up. To this end we substitute the Eq.~\eqref{MF_final_app} into the total momentum of the phonons, $\v P_{B} = \sum_{\v k} \v k \abs{\beta_k}^2$, and obtain the  expression
\begin{eqnarray}\label{PPhononSelfCon}
\v P_{B} &=& \left(\frac{2\pi \mu_{\rm red} ^{-1}\sqrt{n}}{a_{IB}^{-1}-a_+^{-1}}\right)^2   \sum_k  \frac{\vec k W_{\v k}^2}{\left( \omega_{\v k} +\frac{{\v k}^2}{2M} - \frac{\v k}{M} (\v P- \v P_{B})\right)^2}.
\end{eqnarray}
The self-consistent solution of  Eqs.~\eqref{a0_app}  and~\eqref{PPhononSelfCon} allows to find the saddle point for given initial conditions, $a_{IB}$ and $\v P$. 
Finally, the energy is obtained as expectation value of the Hamiltonian~\eqref{eq:H2} in this state. After some algebra one obtains  the expression for energy 
\begin{eqnarray} \nonumber
E_{\rm pol} (\v P)&=&  \frac{{\v P}^2-{\v P}_{B}^2}{2M}  + \frac{2\pi}{\mu_{\rm red}}\frac{n}{a_{IB}^{-1} - a_0^{-1}}.
\end{eqnarray}

\begin{figure*}[t]
\centering
\includegraphics[width=\textwidth]{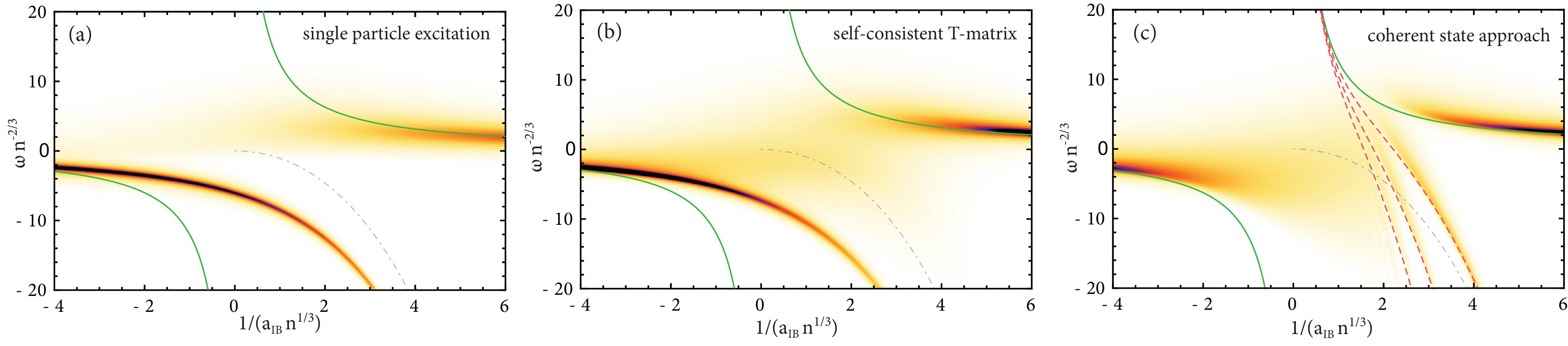}
\vspace{0mm}
\caption{
Comparison of the polaron excitation spectrum as function of frequency and inverse interaction strength as obtained from various theoretical approaches. (a) Time dependent single-particle excitation ansatz (which is equivalent to a non-selfconsistent T-matrix approach previously obtained in \cite{Rath2013}); (b) self-consistent T-matrix approach; (c) time-dependent mean-field/coherent state approach as described in the main body of the text. The results are obtained for a Bose gas with $a_{BB}=0$ and the range of the impurity-bath interaction is set by the UV cutoff scale $n^{-1/3}\Lambda=20$. As in Fig.~(1) of the main text the green line shows the saddle point prediction Eq.~\eqref{mfenergy} for the polaron energy while the gray lines give the two-body dimer binding energy. Finally the red dashed lines in (c) show the energies of the high-order phonon-impurity bound states.}
\label{fig:Chevy}
\end{figure*}

\section{III. Derivation of the energy of the bound states}
In this section we provide the derivation of the bound state energy. We propose the wave-function which accounts for a single Bogoliubov excitation above the polaron state $\ket{\Psi_\text{pol}}$
\begin{equation}\label{boundst}
\ket{\Psi'(t)}=\sum_\veck \gamma_\veck(t) \bed_\veck \ket{\Psi_\text{pol}}.
\end{equation} 
The polaron state in the Lee-Low-Pines frame, i.e.~after the transformation with $\hat S$, can be represented as a displacement operator $\hat U_\text{pol}^\dag \left[\beta_{\v k}\right] = \exp\left( \sum_{\v k} \beta_{\v k } \hat b_{\v k}^\dag - \text{h.c.}\right)$ acting on the vacuum state,
$\ket{\Psi_\text{pol}} = \hat U_\text{pol}^\dag \left[\beta_{\v k}^\text{pol}\right] \ket{0}$. Here the $\beta_{\v k}^\text{pol}$ are defined by  Eq.~\eqref{MF_final_app} in the previous section of the SM. After some  algebra we can rewrite the state $\ket{\Psi'(t)}$ in the form
\begin{equation}
\ket{\Psi'(t)} \equiv \hat U_\text{pol}^\dag \left[\beta_{\v k}^{pol}\right] \left( \gamma_0(t) +\sum_\veck \gamma_\veck(t) \bed_\veck\right) \ket{0}
\end{equation}
where $\gamma_0(t) = \sum_{\v k} \gamma_{\v k}(t) \beta_{\v k}^\text{pol}$.
We calculate the Lagrangian of the system with respect to this state, $\mathcal{L}=\bra{\Psi'(t)}i\partial_t-\hat{\mathcal H}\ket{\Psi'(t)}$, and obtain the equations of motion for the parameters of the wave-function~\eqref{boundst}
\begin{eqnarray}\label{EOMboundstates}
i\partial_t \gamma_0(t) &=& E_\text{pol} \gamma_0(t), \\
i\partial_t \gamma_{\v k}(t) &=& \left(  E_\text{pol}+\Omega_{\v k}\right) \gamma_{\v k}(t) 
+ \frac{g_\Lambda }{2} \sum_{\v k'} \left( W_{\v k}W_{\v k'} +W_{\v k}^{-1}W_{\v k'}^{-1} \right)\gamma_{\v k'}(t).  \nonumber
\end{eqnarray}
The first equation can be solved trivially assuming $\gamma_0 (t) =0$.
This solution implies that the bound state solution is orthogonal to the polaron state.
We are searching for a solution of these equations of motion such that the parameters $\gamma_{\v k}(t)$ 
of the wave-function are time evolving with  frequency $\omega$, $\gamma_k \propto e^{-i\omega t}$. To this end we transform the second line of Eq.~\eqref{EOMboundstates} to the frequency domain and find
\begin{eqnarray}\label{EOMboundstatesFr}
\gamma_{\v k}(\omega) &=&  
\frac{g_\Lambda }{2} \frac{W_{\v k}}{ \omega-E_\text{pol}-\Omega_{\v k}}\sum_{\v k'} W_{\v k'} \gamma_{\v k'}(\omega)  \\ \nonumber 
&+&
\frac{g_\Lambda }{2} \frac{W_{\v k}^{-1}}{ \omega-E_\text{pol}-\Omega_{\v k}}\sum_{\v k'} W_{\v k'}^{-1} \gamma_{\v k'}(\omega).
\end{eqnarray}
Multiplying by $W_{\v k}$ or $W_{\v k}^{-1}$ and summing over over the momentum we obtain the following matrix equation,
\begin{eqnarray}
&&\arr{1 -\frac{g_\Lambda}{2}\sum_{\v k} \frac{W_{\v k}^2}{\omega - E_\text{pol} - \Omega_{\v k}}}
{-\frac{g_\Lambda}{2}\sum_{\v k} \frac{1}{\omega - E_\text{pol} - \Omega_{\v k}}}
{-\frac{g_\Lambda}{2}\sum_{\v k} \frac{1}{\omega - E_\text{pol} - \Omega_{\v k}}}
{1 -\frac{g_\Lambda}{2}\sum_{\v k} \frac{W_{\v k}^{-2}}{\omega - E_\text{pol} - \Omega_{\v k}}} \nonumber\\
&&\quad\otimes\col {\sum_{\v k} W_{\v k} \gamma_{\v k}(\omega)}  {\sum_{\v k} W_{\v k}^{-1} \gamma_{\v k}(\omega)} = 0.
\end{eqnarray}
This equation has a nontrivial solution if and only if the determinant of the matrix is equal to zero. Neglecting terms that are vanishing in the limit $\Lambda \rightarrow \infty$, $ \sim\Lambda^{-2}$, we obtain the following equation for the eigenenergy $\omega $ of the bound state
\begin{equation}\label{eq:BoundstateEn}
1- \frac{g_\Lambda}{2}\sum_{\v k} \frac{W_{\v k}^2+W_{\v k}^{-2}}{\omega - E_\text{pol} - \Omega_{\v k}} =0.
\end{equation}
Using  Eq.~\eqref{eq:LippSchwing} we recast Eq.~\eqref{eq:BoundstateEn} in the form of Eq.~\eqref{chevyonMFEnergy} in the main text. This simple analysis agrees with the features below the polaron branch in the absorption spectrum as shown in Fig.~\ref{fig:Spectrum}.

\section{IV. Comparison with other approaches}
In this section we compare the results for the spectral function $A(\omega)$ of the Bose polaron in the vicinity of a Feshbach resonance, as described by the model~\eqref{eq:H2}, using different theoretical approaches. While in Fig.~\ref{fig:Chevy}(a) we show $A(\omega)$ as obtained from a single-particle expansion using the time-dependent variational state Eq.~\eqref{ChevyFull}, Fig.~\ref{fig:Chevy}(b) shows the result using a self-consistent field-theoretical T-matrix approach \cite{Rath2013}. Finally the results are compared with Fig.~\ref{fig:Chevy}(c) which shows the result from the time-dependent coherent state approach as described by Eq.~\eqref{eq:WF}. All results are obtained for $a_{BB}=0$ and a cutoff scale $n^{-1/3}\Lambda=20$. 
We note that the result shown in Fig.~\ref{fig:Chevy}(a) has been obtained previously in Ref.~\cite{Rath2013} using a non-self-consistent T-matrix (NSCT) approach which is equivalent to the time-dependent single particle excitation ansatz Eq.~\eqref{ChevyFull} when $a_{BB}=0$. 

The main features in the NSCT/single particle excitation approach are the coherent  polaron excitations at negative and positive energies. In this approach the attractive polaron peak  exists at arbitrary interaction strength, and the excitation follows the mean-field prediction $E_{MF}=-\frac{2\pi}{\mu_\text{red}}n a_{IB}$ (green line) in the limit of small dimensionless interaction parameter $n^{1/3}|a_{IB}|\ll1$.
As the Feshbach resonance is crossed to positive $\ainv$ the attractive polaron peak looses weight. However, in contrast to the coherent state approach [cf.~Fig.~\ref{fig:Chevy}(c)], neither the destruction of the polaron quasiparticle at resonance is captured, nor the formation of the series of bound states for increasing $\ainv$. At positive scattering length $a_{IB}$ the repulsive polaron excitation at positive energies is recovered.  It  follows the saddle point prediction only for very weak effective impurity bath interactions.

Away from the coherent excitations, the NSCT/single-particle excitation approach predicts a spectral gap between the attractive polaron and zero energy, cf.~Fig.~\ref{fig:Chevy}(a). This gap is, however, an artifact of the approximation. For instance, in a self-consistent T-matrix  (SCT) approach, shown in Fig.~\ref{fig:Chevy}(b),  such a gap is absent. In this approach an infinite number of particle excitations is taken into account at the basis of a self-consistent calculation of the in-medium T-matrix and  impurity self-energy (the computational details are described in \cite{Rath2013}). The, compared to the NSCT approach, increased number of bosonic excitations leads to an additional renormalization of the spectrum. As one consequence  the spectral gap, visible in the NSCT approach, is absent and furthermore the repulsive polaron looses additional spectral weight as the Feshbach resonance is approached from the side of negative scattering length. This loss of spectral weight is even more pronounced in the coherent state approach, cf.~Fig.~\ref{fig:Chevy}(c), where multiple boson fluctuations lead to the destruction of the attractive polaron peak. 

In contrast to the coherent state approach, the single-particle excitation and SCT approaches do not describe the formation of the series of bound states. In the case of the single-particle excitation approach this fact is simply explained since only one boson excitation is allowed in the wave function.  The SCT approach, which allows in principle more than one bosonic excitation, yet does not reveal higher order bound states. We attribute this fact to the specific momentum structure generated by the SCT approach which  only allows for fluctuations in the pairing channel of the interaction vertex. The particular correlations captured by this momentum structure  seems to be insufficient for a  description of the multiple bound state formation.  Finally we note that both the SCT and the coherent state approach  show a more pronounced repulsive polaron peak as compared to the single-particle excitation approach which can be tested by radio-frequency spectroscopy of impurities immersed in a BEC of ultracold atoms.

\end{document}